# Is Fickian Yet Non-Gaussian Diffusion Ubiquitous?

Alejandro Cuetos[1], Neftalí Morillo[1] and Alessandro Patti[2,*]

[1]*Department of Physical, Chemical and Natural Systems,*
*Pablo de Olavide University, 41013 Sevilla, Spain* and
[2]*School of Chemical Engineering and Analytical Science,*
*The University of Manchester, Manchester, M13 9PL, UK*

Recent studies unveiled the Fickian yet non-Gaussian (FNG) dynamics of many soft matter systems and suggested this phenomenon as a general characteristic of the diffusion in complex fluids. In particular, it was shown that the distribution of particle displacements in Fickian diffusion is not necessarily Gaussian, and thus the Einstein and Smoluchowski theory describing the Brownian motion of individual objects in a fluid would not be applicable. In this Letter, we investigate whether the FNG dynamics so far reported in gels, granular materials, biological and active matter systems, is also a distinctive feature of colloidal liquid crystals. To this end, we perform Brownian Dynamics simulations of oblate and prolate colloidal particles in the nematic phase. We detect a normal and Gaussian dynamics at short and long time scales, whereas, at intermediate time scales, a non-Fickian and non-Gaussian dynamics is found. Additionally, we revisit the nature of the decay of the self-van Hove correlation function, $G_s(r,t)$, which is here approximated with an ellipsoidal, rather than spherical, Gaussian distribution. The new expression that we propose is able to correctly assess the Gaussian dynamics in inherently anisotropic systems, like liquid crystals, where the standard Gaussian approximation of $G_s(r,t)$ would fail.

In 1827, the Scottish botanist Robert Brown reported on pollen grains suspended in water and moving as persistently perturbed by random forces of uncertain nature [1]. Almost eighty years later, in his *annus mirabilis*, Einstein realised that this intriguing, jittery movement, referred to as Brownian motion, was due to the thermal energy that colloidal particles dissipate as a result of their collisions with the surrounding solvent molecules [2]. Einstein's theoretical intuitions, along with the almost simultaneous work by Smoluchowski [3], were fully corroborated experimentally by Perrin in 1909 [4]. In a nutshell, the theory posits that the particle mean-square displacement (MSD) is a linear function of the time $t$ (Fickian diffusion), whereas the particle displacements are Gaussian distributed.

Systems deviating from this behaviour and exhibiting anomalous diffusion, where MSD $\propto t^\alpha$ with $\alpha < 1$ for sub-diffusion and $\alpha > 1$ for super-diffusion, are regularly found for both Gaussian and non-Gaussian distribution of particle displacements [5]. The recent novelty arises from of a wide spectrum of colloidal systems, such as RNA-proteins in cellular cytoplasm [6], tracers in crowded suspensions of colloidal spheres [7], or lipid vesicles in solutions of F-actin filaments [8], that have been found to display Fickian diffusion ($\alpha = 1$), but non-Gaussian distribution of displacements. In some of these systems, at long time scales, short particle displacements follow a Gaussian distribution, but longer displacements do not and, by contrast, are exponentially distributed. This Fickian yet non-Gaussian (FNG) behaviour has been described as the result of the superposition of many Gaussian independent diffusive processes [8]. The observation of FNG diffusion even in systems that, due to their relative simplicity and dilute concentration of particles, would be assumed to follow a canonical Brownian diffusion, has been interpreted as a convincing argument supporting its ubiquitous nature [9]. However, the FNG signature in especially complex systems, such as colloidal liquid crystals (LCs), where an anisotropic diffusion is observed, is still to be explored.

In this Letter, we show that, although many biological, soft and active matter systems display an intriguing and well-documented FNG dynamics, this behaviour is not ubiquitous. By Brownian Dynamics (BD) simulations, we study the dynamics of nematic colloidal LCs of disk-like and rod-like particles and show that a typical Brownian diffusion is observed at short and long time scales. To this end, we first recall the three main time regimes of diffusion in a colloidal suspension. At short time scales, particles diffuse through the solvent and dissipate their thermal energy as a result of the collisions with the solvent molecules. This regime is diffusive (or Fickian) and the MSD is a linear function of time, or $\langle \Delta r^2 \rangle \propto t$, as reported for both oblate and prolate particles in the Supplemental Material [10]. In particular, $\langle \Delta r^2 \rangle = 2dD_s t$, with $d$ the dimensionality of the move and $D_s$ the translational diffusion coefficient of an isolated particle in a medium. At intermediate time scales, the diffusion of individual particles is slowed down by a sort of temporary cage formed by other particles [11]. The duration of this caging effect is mainly determined by the system packing and interparticle interactions. Fi-nally, at long time scales, the diffusion is controlled by the inter-particle collisions and the Fickian regime is recovered. In this case, $\langle \Delta r^2 \rangle = 2dD_l t$, with $D_l$ the long-time translational diffusion coefficient, being in general smaller than its short-time counterpart [12].

*electronic address: alessandro.patti@manchester.ac.uk

At each of this time scales, the distribution of the particle displacement can be measured by the self-part of the van Hove correlation function, $G_s(r,t) = 1/N \left\langle \sum_{i=1}^{N} \delta(r - |\mathbf{r}_i(t) - \mathbf{r}_i(0)|) \right\rangle$, where $N$ is the total number of particles, $\delta$ the Dirac-delta, and $\langle ... \rangle$ denotes ensemble average over different trajectories. If the displacements are Gaussian distributed, then $G_s(r,t)$ is a Gaussian function of $r$ at all times:

$$G_{s,d}(r,t) = (4\pi D_t t)^{-d/2} \exp\left(-\frac{r^2}{4D_t t}\right), \quad (1)$$

where the subindex $t$ indicates a generic dependence on time of the diffusion coefficient, such that $D_t = D_s$ and $D_t = D_l$ at short and long time scales, respectively. The difficulty to determine the time regimes in which a Gaussian diffusion actually holds and perform measurements at very large length and long time scales, a limit where the non-Gaussian character of Fickian diffusion is especially challenging to be proven [14], has challenged the general applicability of Eq. 1 to complex fluids [13].

Investigating the diffusion in nematic LCs is particularly attractive because of the anisotropic nature of the particles' dynamics and the opportunity to address its eventual FNG character over two independent directions. In addition, the relatively moderate packing fractions of nematics, as compared to smectic or columnar LC phases, allow to more easily achieve the asymptotic limit of long time scales and distinguish it from the non-Gaussian signature of the caging effect at shorter times. Previous works observed the existence of deviations of $G_s(r,t)$ from a Gaussian distribution in smectic [15–17] and columnar [18, 19] LCs, where the diffusion perpendicular, respectively, to the layers and columns is especially slow and determining the onset of the long time diffusive regime not always straightforward.

The nematic LCs studied here consist of monodisperse rods or disks. Rods are modelled as prolate spherocylinders with aspect ratio $a_p = (L + \sigma)/\sigma$, where $L$ and $\sigma$ are the length and diameter, respectively, of a cylinder capped by two hemispheres of diameter $\sigma$. Disks are modelled as oblate spherocylinders of aspect ratio $a_o = a_p$, consisting of a cylindrical body of height $\sigma$ and diameter $L$ and surrounded by a toroidal rim with tube radius $\sigma/2$. The aspect ratio of the two sets of particles is set to 15.6. In both cases, the interparticle interactions are described by the Soft Repulsive Spherocylindrical (SRS) potential, employed in the past to investigate the phase behaviour of prolate and oblate particles [20, 21].

Temperature in both systems is set to $T^* = k_B T/\epsilon = 1.465$, with $k_B$ the Boltzmann constant and $\epsilon$ the strength of the SRS potential. At this temperature, nematic LCs are observed at packing fractions $\eta = 0.35$ for both prolate and oblate particles, where $\eta = Nv/V$, $v$ is the particle volume, and $V$ the volume of the simulation box. All the simulations have been run at this packing fraction, although few other packings have been explored to check the robustness of our conclusions. Our system's length unit is $\sigma$, while the time unit is $\tau = \sigma^3 \mu / k_B T$, where $\mu$ is the solvent viscosity. For additional details on the model and BD simulation methodology, the interested reader is referred to the Supplemental Material [10] and former works [22, 23]. Here, we briefly remind that the short-time diffusion coefficients employed to update the particle position and orientation are those calculated by Shimizu for prolate and oblate spheroids [24]. In particular, $D_{s,\parallel}$ and $D_{s,\perp}$ are the diffusion coefficients parallel and perpendicular to the main particle axis, respectively, while $D_{s,\theta}$ is the rotational diffusion coefficient. For prolate particles $D_{s,\parallel} > D_{s,\perp}$, while for oblate particles $D_{s,\parallel} < D_{s,\perp}$. As previously reported, the long-time isotropic diffusion of these systems is Fickian, with $D_l = \lim_{t\to\infty} \langle \Delta r^2 \rangle / 6t$, as well as the diffusion parallel and perpendicular to the nematic director $\hat{\mathbf{n}}$, with $D_{l,\parallel} = \lim_{t\to\infty} \langle \Delta r_\parallel^2 \rangle / 2t$ and $D_{l,\perp} = \lim_{t\to\infty} \langle \Delta r_\perp^2 \rangle / 4t$ [22, 25, 26]. These results agree very well with our calculation of the MSD for both sets of particles shown in the Supplemental Material, where we also detect a non-Fickian (sub-diffusive) regime at intermediate time scales [10].

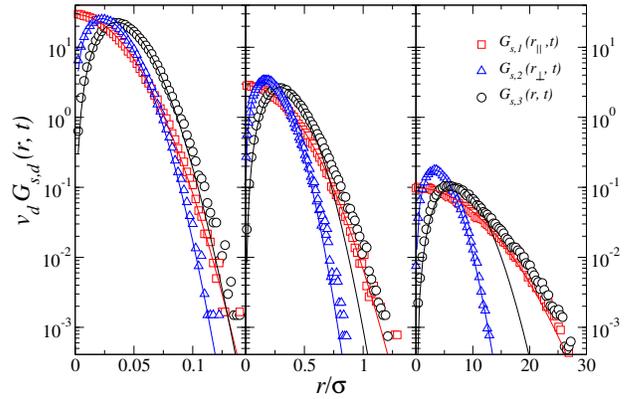

FIG. 1: Parallel ($\square$), perpendicular ($\triangle$) and total ($\circ$) self-van Hove correlation functions for a nematic LC of prolate particles at times $t/\tau = 10^{-2}$ (left frame), $t/\tau = 1$ (middle frame) and $t/\tau = 10^3$ (right frame). Results are normalised by $v_d = 1$, $2\pi r_\perp$ or $4\pi r^2$ for $d = 1$, 2 or 3, respectively. Symbols are simulation results, while solid lines are Gaussian approximations as given in Eq. 1, with $D_t$ fitting parameter.

In the light of these preliminary considerations, which highlight the Fickian nature of the short-time and long-time diffusion, we now consider whether the particle displacements as well as their parallel and perpendicu-lar projections to $\hat{\mathbf{n}}$ are Gaussian distributed. To this end, we calculate the parallel, $G_{s,1}(r_\parallel,t)$, perpendicular, $G_{s,2}(r_\perp,t)$, and total, $G_{s,3}(r,t)$, self-van Hove correlation functions, which are shown in Figs. 1 and 2 for rods and disks, respectively, at short (left frames), intermediate (middle frames), and long (right frames) time scales. In the same figures, we fit our simulation results with the Gaussian approximations calculated from Eq. 1, where $D_t$ is a fitting parameter.

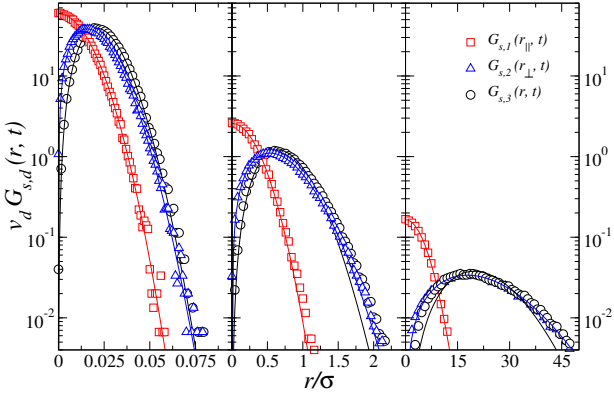

FIG. 2: Parallel ($\square$), perpendicular ($\triangle$) and total ($\circ$) self-van Hove correlation functions for a nematic LC of oblate particles at times $t/\tau = 10^{-2}$ (left frame), $t/\tau = 10.8$ (middle frame) and $t/\tau = 2 \times 10^4$ (right frame). Results are normalised by $v_d = 1$, $2\pi r_\perp$ or $4\pi r^2$ for $d = 1$, 2 or 3, respectively. Symbols are simulation results, while solid lines are Gaussian approximations as given in Eq. 1, with $D_t$ fitting parameter.

We observe that the Gaussian approximation to $G_{s,1}(r_\parallel,t)$ and $G_{s,2}(r_\perp,t)$ is very good at both short and long time scales, while, at intermediate times, where the diffusion is however not Fickian, moderate discrepancies are detected. We then conclude that, at least in the direction of $\hat{\mathbf{n}}$ and perpendicularly to it, prolate and oblate colloidal particles exhibit Fickian and Gaussian diffusion at short times. A more detailed analysis deserves the total self-van Hove function, $G_{s,3}(r,t)$, which clearly appears underestimated by the Gaussian fit at intermediate and long time scales, and less significantly also at short times (circles and black lines in Figs. 1 and 2). While at intermediate times the diffusion is not Fickian and a non-Gaussian behaviour is not especially surprising, at short and long times one would conclude that prolate and oblate particles follow an FNG diffusion. Nevertheless, we notice that the Gaussian approximation in Eq. 1 results from the integration of the Langevin equation under the assumption of a spatial isotropy, where $D_{s,\parallel} = D_{s,\perp}$ and $D_{l,\parallel} = D_{l,\perp}$ [12]. This assumption does not hold in a nematic LC and, more generally, in any complex fluid with anisotropic morphology.

Therefore, we propose an ellipsoidal, rather than spherical, Gaussian approximation of $G_{s,3}(r,t)$, where the displacements in the direction parallel and perpendicular to $\hat{\mathbf{n}}$ are still assumed to be Gaussian distributed, but independent from each other [27]. The new form of the total self-van Hove correlation function is determined by combining the displacements' distributions along the parallel and perpendicular directions to $\hat{\mathbf{n}}$ and reads

$$G_s(r_\parallel, r_\perp, t) = G_{s,1}(r_\parallel, t)\, G_{s,2}(r_\perp, t) = \frac{1}{\left((4\pi t)^3 D_{t,\perp}^2 D_{t,\parallel}\right)^{1/2}} \exp\left(-\frac{r_\parallel^2}{4D_{t,\parallel} t} - \frac{r_\perp^2}{4D_{t,\perp} t}\right) \quad (2)$$

where $G_{s,1}(r_\parallel,t)$ and $G_{s,2}(r_\perp,t)$ have been obtained substituting, respectively, $d = 1$ and $d = 2$ in Eq. 1. The probability to find a particle at distance $r = (r_\perp^2 + r_\parallel^2)^{1/2}$ from its original position at $t = 0$, is obtained by integrating the above equation over a spherical surface of radius $r$:

$$G'_{s,3}(r,t) = \frac{\int_S dS\, G_s(r_\parallel, r_\perp, t)}{\int_S dS} \quad (3)$$

The solution of the above integral can either take the form

$$G'_{s,3}(r,t) = \frac{\Omega}{(4\pi t)^{3/2}} \exp\left(-\frac{r^2}{4D_{t,\parallel} t}\right) \frac{F(r\Delta_p^{1/2})}{r\Delta_p^{1/2}} \quad (4)$$

or equivalently

$$G'_{s,3}(r,t) = \frac{\Omega\sqrt{\pi}}{2(4\pi t)^{3/2}} \exp\left(-\frac{r^2}{4D_{t,\perp} t}\right) \frac{\mathrm{erf}(r\Delta_o^{1/2})}{r\Delta_o^{1/2}}, \quad (5)$$

where $F(...)$ is the Dawson's intergral, $\mathrm{erf}(...)$ the error function, $\Omega = 1/(D_{t,\perp}^2 D_{t,\parallel})^{1/2}$, and $\Delta_p = -\Delta_o = 1/(4D_{t,\perp}) - 1/(4D_{t,\parallel} t)$. Eqs. 4 and 5 are mathematically identical, being the former more suitable for prolate geometries, where $D_{t,\parallel} > D_{t,\perp}$, and the latter for oblate geometries, where $D_{t,\parallel} < D_{t,\perp}$.

The total self-van Hove functions calculated from Eqs. 4 and 5 are shown as solid lines in Fig. 3, along with our simulation results. For comparison, we also show the Gaussian approximation of $G_{s,3}(r,t)$ as obtained from Eq. 1 (dashed lines). The agreement between simulations and theoretical predictions is excellent, confirming the Gaussian nature of the Fickian diffusion at long times and thus discarding the occurrence of an FNG diffusion for the two particle geometries. We stress that the dashed and solid curves in Fig. 3 are not fits, as the diffusion coefficients, $D_{l,\parallel}$ and $D_{l,\perp}$, at long times have been obtained from the corresponding MSDs. The theoretical predictions of $G'_{s,1}(r_\perp,t)$ and $G'_{s,2}(r_\parallel,t)$ are also in excellent agreement with the simulation results and are included in the Supplemental Material [10].

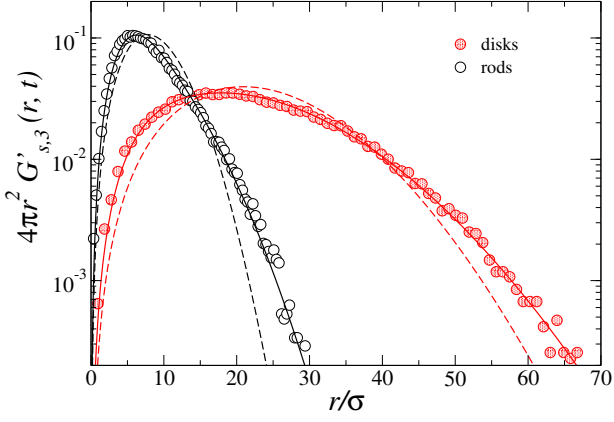
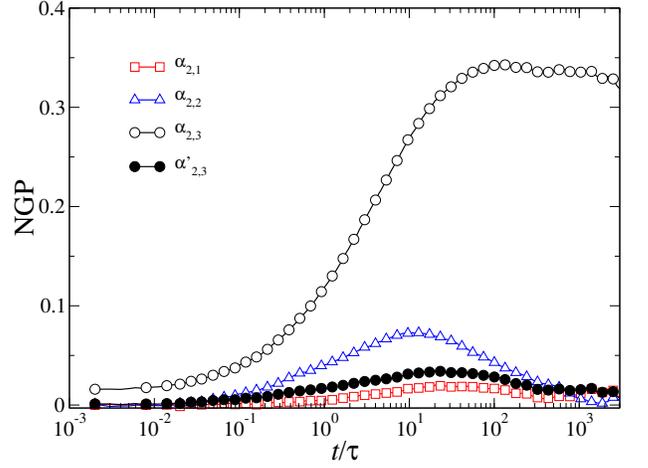

FIG. 3: Total self-van Hove correlation functions for a nematic phase of rod-like and disk-like particles at $t/\tau = 10^3$ and $2 \times 10^4$, respectively. Symbols are simulation results, dashed lines are Gaussian distributions obtained from Eq. 1, and black solid lines are Gaussian distributions obtained with Eq. 4 (rods) and 5 (disks).

FIG. 4: Parallel ($\square$), perpendicular ($\triangle$) and total ($\circ$) non-Gaussian parameters for a nematic fluid of prolate particles as calculated from Eq. 6. The solid circles ($\bullet$) indicate the total NGP calculated from Eq. 10. Symbols are simulation results and solid lines are guides for the eye.

The non-Gaussian character of $G_{s,d}(r,t)$ can also be assessed by expanding this function in a series of Hermite polynomials, whose first term incoporates most of the function's non-Gaussianity in the following coefficient [28]:

$$\alpha_{2,d}(t) = \frac{\langle \Delta r^4(t) \rangle}{(1+2/d)\langle \Delta r^2(t)\rangle^2} - 1. \quad (6)$$

In particular, $\alpha_{2,d}(t)$, which is usually referred to as non-Gaussian parameter (NGP), vanishes if no deviations from Gaussian behaviour are observed. Parallel ($d=1$), perpendicular ($d=2$), and total ($d=3$) NGPs are plotted in Fig. 4 for nematic phases of prolate particles. Very similar results are observed for oblate particles and are not shown here. At short time scales, $\alpha_{2,1}$ and $\alpha_{2,2}$ are very close to zero, whereas $\alpha_{2,3}$ is clearly different than zero even at very short times. At intermediate times, when the diffusion is not Fickian, both parallel and perpendicular NGPs are observed to increase, but this tendency is especially evident for the latter, whose maximum value is achieved at approximately $t/\tau = 12$. We notice that the total NGP predicted by Eq. 6, $\alpha_{2,3}$, significantly increases in this time regime. At long times, both the parallel and perpendicular NGPs start to decrease, reaching values very close to zero. Different is the tendency displayed by $\alpha_{2,3}$, which seems to reach a maximum at roughly $t/\tau = 10^2$ and then eventually decays over a time scale that goes beyond our simulation time. This result is however obtained by employing a Gaussian form of the self-van Hove function that is not able to describe the dynamics of anisotropic systems. By following similar arguments to those illustrated above, we employ Eqs. 4 and 5 to derive an expression for the total NGP that incoporates parallel and perpendicular diffusion coefficients. To this end, we first re-write the total NGP as

$$\alpha_{2,d}(t) = K\left[\langle \Delta r^4 \rangle / \langle \Delta r^2 \rangle^2\right]_{\text{sim}} - 1, \quad (7)$$

where $K \equiv \left[\langle \Delta r^2 \rangle^2/\langle \Delta r^4 \rangle\right]_{\text{th}}$ and the functions in $[\cdots]_{\text{sim}}$ and $[\cdots]_{\text{th}}$ are calculated, respectively, by simulation and employing the theoretical distribution of the displacements. More specifically, if we make use of the Gaussian distribution given in Eq. 1, then $K$ takes the values $1/3$, $1/2$ or $3/5$, for $d = 1$, 2 or 3, respectively, and the standard form of the NGP (Eq. 6) is recovered. Alternatively, if we incorporate the space anisotropy by using Eqs. 4 and 5, the theoretical values of $\langle \Delta r^2 \rangle$ and $\langle \Delta r^4 \rangle$ read

$$\langle \Delta r^2 \rangle = \int_0^\infty r^4 G_{s,3}(r,t) \, dr = \left(2D_{t,\parallel} + 4D_{t,\perp}\right) t \quad (8)$$

and

$$\langle \Delta r^4 \rangle = \int_0^\infty r^6 G_{s,3}(r,t) \, dr = \\ 4\left(3D_{t,\parallel}^2 + 8D_{t,\perp}^2 + 4D_{t,\parallel}D_{t,\perp}\right) t^2 \quad (9)$$

We can now define an alternative form of the NGP, which reads

$$\alpha'_{2,3} = \frac{D_{t,\parallel}^2 + 4D_{t,\perp}^2 + 4D_{t,\parallel}D_{t,\perp}}{3D_{t,\parallel}^2 + 8D_{t,\perp}^2 + 4D_{t,\parallel}D_{t,\perp}} \frac{\langle \Delta r^4 \rangle}{\langle \Delta r^2 \rangle^2} - 1 \quad (10)$$

Similarly to the NGP given in Eq. 6, also $\alpha'_{2,3}$ can be applied to any particle geometry. The key difference is that $\alpha'_{2,3}$ depends on the instantaneous value of the diffusion coefficients parallel and perpendicular to $\hat{\mathbf{n}}$, as highlighted by the subindex $t$ in the equations above. In particular, to calculate $\alpha'_{2,3}$ over time, the diffusivities $D_{t,\parallel}$ and $D_{t,\perp}$ have been estimated from the instantaneous values of the MSD as obtained by computer simulation. We plot $\alpha'_{2,3}$ in Fig. 4, where it is compared to the total NGP, $\alpha_{2,3}$, that has been derived neglecting the anisotropy of diffusion. As already found for $G'_{s,3}(r,t)$, the diffusion at short time scales appears to be Gaussian, with $\alpha'_{2,3} \approx 0$ for $t/\tau < 10^{-1}$. At intermediate times, $\alpha'_{2,3}$ becomes slightly larger than zero, revealing deviations from Gaussian behaviour, which are anyway significantly softer than those detected with $\alpha_{2,3}$ and consistent with those of $\alpha_{2,1}$ and $\alpha_{2,2}$. At $t/\tau > 10^2$, when the diffusion recovers its Fickian nature [10], $\alpha'_{2,3}$ reaches again values that are very close to zero.

In summary, we conclude that FNG diffusion is not ubiquitous in soft matter and the mathematical tools to properly assess it should not *a priori* neglect the impact of space anisotropy. More specifically, our results show that colloidal particles in nematic LCs display a Fickian and Gaussian dynamics at short and sufficiently long time scales, while at intermediate times, when the particles experience a caging effect imposed by their neighbours, the diffusion is sub-diffusive and non-Gaussian. We have shown that the Fickian and Gaussian dynamics of colloidal nematic LCs cannot be appreciated by a distribution function of particle displacements that assumes space symmetry and calculated via the standard self-van Hove correlation function. To overcome this limitation, we propose an ellipsoidal Gaussian distribution that takes into account the diffusion coefficients parallel and perpendicular to the nematic director. This new distribution function is able to reproduce our simulation results with remarkable precision and is crucial to understand the nature of the diffusion in colloidal LCs, which does not show evidence of an FNG signature. The new form of the self-van Hove functions is applied to formulate a non-Gaussian parameter that incoporates the instantaneous value of the diffusion coefficients and is able to quantify deviations from Gaussian behaviour more precisely. Our theoretical formalism is relevant to assess the existence of Gaussian dynamics in a number of anisotropic systems, including for instance crystalline porous materials as zeolites.

AC and NM acknowledge project P12-FQM-2310 funded by the Junta de Andalucía-FEDER and C3UPO for HPC facilities provided. AC also acknowledges grant PPI1719 awarded by the Pablo de Olavide University for funding his research visit to the School of Chemical Engineering and Analytical Science, The University of Manchester. AP acknowledges financial support from EPSRC under grant agreement EP/N02690X/1. All authors wish to thank Gonzalo Angulo (Institute of Physical Chemistry of the Polish Academy of Sciences) for insightful discussions.

---

# Is Fickian Yet Non-Gaussian Diffusion Ubiquitous?
# Supplemental Material

Alejandro Cuetos[1], Neftalí Morillo[1] and Alessandro Patti[2,*]
[1]*Department of Physical, Chemical and Natural Systems,*
*Pablo de Olavide University, 41013 Sevilla, Spain and*
[2]*School of Chemical Engineering and Analytical Science,*
*The University of Manchester, Manchester, M13 9PL, UK*

## I. MODEL AND SIMULATION METHODS

We have modelled both prolate (rods) and oblate (disks) particles as spherocylinders. A spherocylinder is a solid of revolution obtained by rotating a rectangle of length $L$ capped by two semicircles of diameter $\sigma$ at both ends. If this 2D shape rotates around the segment connecting the centres of the two semicircles, the resulting solid is a prolate spherocylinder, consisting of a cylindrical body capped by two semispheres. By contrast, if it rotates around the axis perpendicular to this segment, the solid generated is an oblate spherocylinder, consisting of a cylindrical body surrounded by a toroidal rim. Therefore, the shape anisotropy for oblate and prolate particles is defined as $a_p = a_o = (L + \sigma)/\sigma$ [1].

All particles interact via the Soft Repulsive Spherocylinder potential (SRS), being obtained by truncating and shifting the 12-6 Kihara potential [2, 3]:

$$U_{SRS} = \begin{cases} 4\epsilon \left[ (\sigma/d_m)^{12} - (\sigma/d_m)^6 + 1/4 \right] & d_m \leq \sqrt[6]{2}\,\sigma \\ 0 & d_m > \sqrt[6]{2}\,\sigma \end{cases}$$

Here $\sigma$ is the diameter of the cylinder in case of rods, and the thickness in case of disks, whereas $d_m$ is the minimum distance between the central cores of the particles, a segment of elongation $L$ for prolate particles, and a disk of diameter $L$ for oblates. Efficient algorithms to calculate the minimum distance for both geometries have been published previously [4, 5].

To simulate the Brownian motion of the particles, we have carried out Brownian Dynamics (BD) simulations, where the position and orientation of each particle $j$ over time are determined by the following set of equations [6]:

$$\mathbf{r}_j^{\parallel}(t + \Delta t) = \mathbf{r}_j^{\parallel}(t) + \frac{D_{\parallel}}{k_B T}\mathbf{F}_j^{\parallel}(t)\Delta t + (2D_{\parallel}\Delta t)^{1/2} R^{\parallel}\hat{\mathbf{u}}(t) \qquad (1)$$

$$\mathbf{r}_j^{\perp}(t + \Delta t) = \mathbf{r}_j^{\perp}(t) + \frac{D_{\perp}}{k_B T}\mathbf{F}_j^{\perp}(t)\Delta t + (2D_{\perp}\Delta t)^{1/2} \left( R_1^{\perp}\hat{\mathbf{v}}_{j,1}(t) + R_2^{\perp}\hat{\mathbf{v}}_{j,2}(t) \right) \qquad (2)$$

$$\hat{\mathbf{u}}_j(t + \Delta t) = \hat{\mathbf{u}}_j(t) + \frac{D_{\vartheta}}{k_B T}\mathbf{T}(t) \times \hat{\mathbf{u}}(t)\Delta t + (2D_{\vartheta}\Delta t)^{1/2} \left( R_1^{\vartheta}\hat{\mathbf{v}}_{j,1}(t) + R_2^{\vartheta}\hat{\mathbf{v}}_{j,2}(t) \right) \qquad (3)$$

where $\mathbf{r}_j^{\parallel}$ and $\mathbf{r}_j^{\perp}$ are the projections of the position vector $\mathbf{r}$ on the direction parallel and perpendicular to the unit vector $\hat{\mathbf{u}}_j$, respectively; $\mathbf{T}_j$ is the total torque acting over particle $j$ [7], $\mathbf{F}_j^{\parallel}$ and $\mathbf{F}_j^{\perp}$ are the components of the forces, respectively, parallel and perpendicular to $\hat{\mathbf{u}}_j$; $R^{\parallel}$, $R_1^{\perp}$, $R_2^{\perp}$, $R_1^{\vartheta}$ and $R_2^{\vartheta}$ are independent Gaussian random numbers of variance 1 and zero mean; and $\hat{\mathbf{v}}_{j,1}$ and $\hat{\mathbf{v}}_{j,1}$ are two random unit vectors, perpendicular to each other and to $\hat{\mathbf{u}}_j$.

The diffusion coefficients at infinite dilution, $D_{\parallel}$, $D_{\perp}$ and $D_{\vartheta}$, have been calculated both for prolate and oblate particles with the analytical expressions proposed by Shimizu for spheroids [8]:

---

* electronic address: alessandro.patti@manchester.ac.uk

$$D_\perp = D_0 \frac{(2a^2 - 3b^2)S + 2a}{16\pi(a^2 - b^2)}b,$$

$$D_\parallel = D_0 \frac{(2a^2 - b^2)S - 2a}{8\pi(a^2 - b^2)}b, \quad (4)$$

$$D_\vartheta = 3D_0 \frac{(2a^2 - b^2)S - 2a}{16\pi(a^4 - b^4)}b,$$

With $D_0 = k_B T/\mu\sigma$, $k_B$ the Boltzmann constant, $T$ the temperature, and $\mu$ the solvent viscosity. $S$ is a geometrical parameter that for prolate particles is given by

$$\text{with} \quad S = \frac{2}{(a^2 - b^2)^{1/2}} \log \frac{a + (a^2 - b^2)^{1/2}}{b}, \quad (5)$$
$$(a = (L+\sigma)/2, \ b = \sigma/2)$$

while for oblate particles is calculated as

$$\text{with} \quad S = \frac{2}{(a^2 - b^2)^{1/2}} \arctan \frac{(b^2 - a^2)^{1/2}}{a}, \quad (6)$$
$$(a = \sigma/2, \ b = (L+\sigma)/2)$$

Simulations have been performed in cubic boxes containing $N = 1344$ rods or $1500$ disks, and packing fraction $\eta = 0.35$, with $\eta = v_m\rho$, $\rho$ the particle density and $v_m$ the volume of the particles [1]. The time step has been set in the range $10^{-4} < t/\tau < 2 \cdot 10^{-3}$, with $\tau = \sigma^3\mu/k_BT$.

The most relevant observables in this work are the parallel, perpendicular and total self-van Hove functions, which are calculated, respectively, as

$$G_{s,1}(r_\parallel, t) = \frac{1}{N}\left\langle \sum_{i=1}^N \delta(r_\parallel - |\mathbf{r}_{\parallel,i}(t) - \mathbf{r}_{\parallel,i}(0)|) \right\rangle \quad (7)$$

$$G_{s,2}(r_\perp, t) = \frac{1}{N}\left\langle \sum_{i=1}^N \delta(r_\perp - |\mathbf{r}_{\perp,i}(t) - \mathbf{r}_{\perp,i}(0)|) \right\rangle \quad (8)$$

$$G_{s,3}(r, t) = \frac{1}{N}\left\langle \sum_{i=1}^N \delta(r - |\mathbf{r}_i(t) - \mathbf{r}_i(t)|) \right\rangle \quad (9)$$

where the symbol $\delta$ is the Dirac delta, the angular brackets denote ensemble average over at least 100 different trajectories and all the particles, and $r_\parallel$ and $r_\perp$ are, respectively, the projections of the displacement parallel and perpendicular to the nematic director $\hat{\mathbf{n}}$. The director $\hat{\mathbf{n}}$ is calculated with the standard procedure of diagonalization of the traceless tensor incoporating the particles' orientation vectors [9]. The functions in Eqs. 7, 8, and 9 should be normalised as follows: $\int_0^\infty G_{s,1} dr_\parallel = \int_0^\infty 2\pi r_\perp G_{s,2} dr_\perp = \int_0^\infty 4\pi r^2 G_{s,3} dr = 1$.

We have also calculated the mean square displacement, defined as

$$\langle \Delta r^2(t) \rangle = \left\langle \frac{1}{N} \sum_{j=1}^N (\mathbf{r}_j(t) - \mathbf{r}_j(0))^2 \right\rangle. \quad (10)$$

The parallel and perpendicular projections of the mean square displacement to $\hat{\mathbf{n}}$ have also been calculated.

## II. FICKIAN BEHAVIOR OF THE MEAN SQUARE DISPLACEMENT

In Figs. 1 and 2, we plot the total (circles), parallel (squares) and perpendicular (triangles) mean square displacement for prolate and oblate particles, respectively. At short and long time scales, they have been fitted to linear functions indicated by solid (short times) and dashes (long times) lines. The quality of these fits confirms the Fickian character of the diffusion at short and long times.

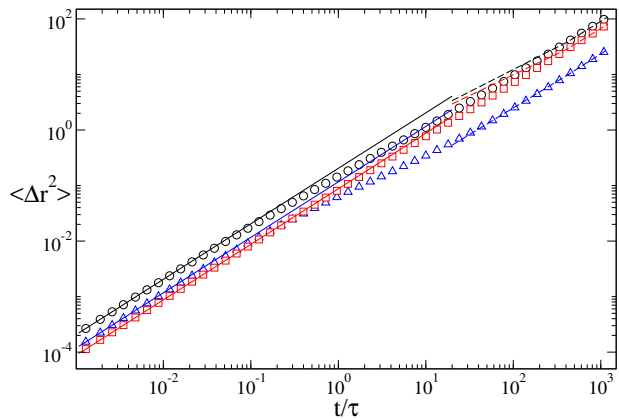

FIG. 1: (Color online). Total (black circles), parallel (red squares) and perpendicular (blue triangles) components of the mean square displacements obtained by computer simulation for prolate particles with anisotropy $a_p = 15.6$ in nematic fluids of packing fraction $\eta = 0.35$. The lines are linear fittings to the simulation results at short (solid lines) and long (dashed lines) times.

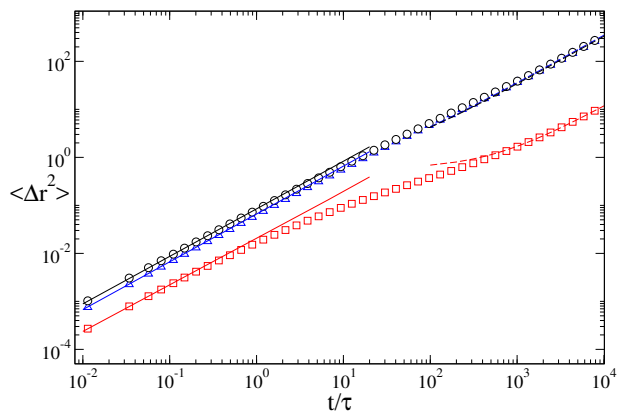

FIG. 2: (Color online). Same as Fig. 1, but for oblates particles with anisotropy $a_o = 15.6$.

## III. DIRECTIONAL SELF-VAN HOVE FUNCTIONS

A comparison between the theoretical predictions of the self-van Hove functions and that calculated by simulations is shown in Fig. 3 for fluid of prolate particles and in Fig. 4 for oblate particles, in both cases at long time scales. The theoretical functions are calculated using Eq. 1 given in the letter, with $d = 1$ for parallel and $d = 2$ for perpendicular diffusion, using the instantaneous values of $D_{l,\parallel}$ and $D_{l,\perp}$ obtained from the mean square displacement.

---

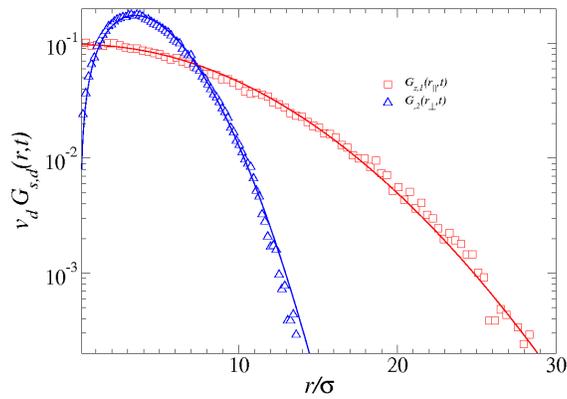

FIG. 3: (Color online). Parallel (red squares and line) and perpendicular (blue triangles and line) self-van Hove correlation functions for a nematic phase of rod-like particles at $t/\tau = 10^3$. Results are normalised by $v_d = 1$, $2\pi r_\perp$ or $4\pi r^2$ for $d = 1, 2$ or 3, respectively. Symbols are simulation results. Lines are Gaussian distributions (see Letter for details).

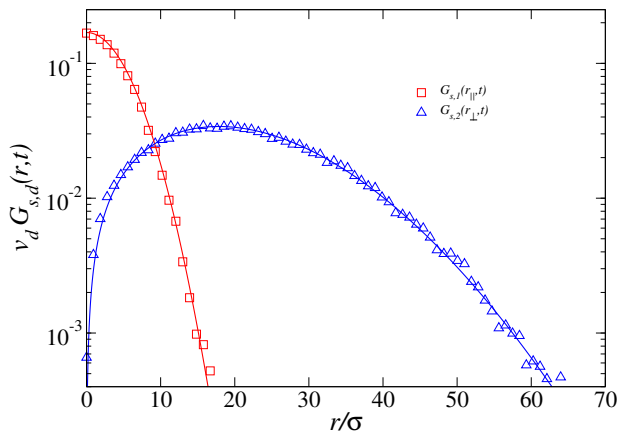

FIG. 4: (Color online). Parallel (red squares and line) and perpendicular (blue triangles and line) self-van Hove correlation functions for a nematic phase of disk-like at $t/\tau = 2 \times 10^4$. Results are normalised by $v_d = 1$, $2\pi r_\perp$ or $4\pi r^2$ for $d = 1, 2$ or 3, respectively. Symbols are simulation results. Lines are Gaussian distributions (see Letter for details).